\documentclass[twocolumn,showpacs,amsmath,amssymb]{revtex4}
\usepackage{psfig}
\usepackage{graphicx}
\begin{document}


\title{ \large\bf Evidence for two-quark content of $f_{0}(980)$ in exclusive $b\to c$ decays }

\author{ \bf Chuan-Hung Chen }

\email{chchen@phys.sinica.edu.tw}

\affiliation{  Institute of Physics, Academia Sinica, Taipei,
Taiwan 115, Republic of China }

\date{\today}

\begin{abstract}
Inspired by a large decay branching  ratio (BR) of $B^{+}\to
f_{0}(980)K^{+}$  measured by Belle recently, we propose that
a significant evidence of the component of
$n\bar{n}=(u\bar{u}+d\bar{d})/\sqrt{2}$ in $f_{0}(980)$ could be
demonstrated in exclusive $b\to c$ decays by the observation of
$f_{0}(980)$ in the final states $\bar{B}\to D^{0(*)} \pi^{+}
\pi^{-}(KK)$ and $\bar{B}\to J/\Psi \pi^{+} \pi^{-}(KK)$. We
predict the BRs of $\bar{B}\to D^{0(*)} (J/\Psi) f_{0}(980)$ to be
${\cal {O}}(10^{-4})$ (${\cal {O}}(10^{-5})$) while the unknown
wave functions of $D^{(*)0}$ ($J/\Psi$) are chosen to fit the
observed decays of $\bar{B}\to D^{(*)0} \pi^{0}\ (J/\Psi K^{0(*)})$.
\end{abstract}

\maketitle


In spite of the successful quark model and QCD theory for strong
interaction, the fundamental questions on the inner structure of
lightest scalar mesons, such as $f_{0}(400-1200)$, $f_{0}(980)$
and $a_{0}(980)$ etc., are still uncertain, even although it has been over
thirty years since $f_{0}(980)$ was discovered first in phase
shift analysis of elastic $\pi \pi$ scattering \cite{f0}. Besides
the interpretations of $qq\bar{q}\bar{q}$ four-quark states
\cite{4q} or $K\bar{K}$ molecular states \cite{KK} or $q \bar{q}$
states \cite{qq} etc., the possibilities of gluonium states
\cite{MO} and scalar glueballs \cite{Robson} are also proposed. It
might be oversimple to regard them as only one kind of
composition.

It is suggested that in terms of $\gamma \gamma$ \cite{MO} and
radiative $\phi$ \cite{Antonelli,BELNP} decays, the nature of
scalar mesons can be disentangled. However, with these
experiments, the conclusions such as given by Refs. \cite{De,DP}
and Ref. \cite{AG} are not unique. The former prefers $q \bar{q}$
while the latter is four-quark content. Nevertheless, according to
the data of E791 \cite{E791} and Focus \cite{FOCUS}, the
productions of scalar mesons which are reconstructed from $D$ and
$D_{s}$ decaying to three-pseudoscalar final states and mainly
show $q \bar{q}$ contents, can provide us a further resolution
\cite{Meadows}. In addition, $Z_{0}$ decay data of OPAL
\cite{OPAL} also hint that $f_{0}(980)$, $f_{2}(1270)$ and
$\phi(1020)$ have the same internal structure. Hence, the
compositions of light scalar bosons should be examined further.


Recently, the decay of $B^{+}\to f_0(980) K^{+}$ with the BR
product of $Br(B^{+}\to f_{0}(980)K^{+})\times Br(f_{0}(980)\to
\pi^+ \pi^-)= (9.6^{+2.5+1.5+3.4}_{-2.3-1.5-0.8})\times 10^{-6}$
has been observed in Belle \cite{Belle1}. The observation not only
displays in the first time $B$ decay to scalar-pseudoscalar final
states but also provides the chance to understand the
characteristics of scalar mesons. Since $B$ meson is much heavier
than $D_{(s)}$ mesons, in the two-body $B$ decays, the outgoing
light mesons will behave as massless particles so that the
perturbative QCD (PQCD) approach \cite{LB,Li}, in which the
corresponding bound states are expanded by Fock states, could
apply. Therefore, as compared to two-parton states, the
contributions of four-parton and gluonium states belong to higher
Fock states. Consequently, we think that the effects of $q \bar{q}$
state are more important than those in $D_{s}$ decays. In this
paper, in order to further understand what  the nature of
$f_{0}(980)$ in $B$ decays is, we take it to be composed of $q
\bar{q}$ states mainly and use $|f_{0}(980)>=\cos\phi_{s} |s
\bar{s}>+\sin\phi_{s} |n \bar{n}>$ with $n
\bar{n}=(u\bar{u}+d\bar{d})/\sqrt{2}$ to denote its flavor wave
function. We note that so far $\phi_{s}$ could be
$42.14^{+5.8^{0}}_{-7.3}$ \cite{MO} and $138^{0}\pm {6^0}$
\cite{AAN}. With the lowest order criterion, the effects of
four-parton and gluonium states are neglected.

Inspired by the large BR of $B^{+}\to f_{0}(980)K^{+}$, we
propose that a significant evidence of the component of $n
\bar{n}$ in $f_{0}(980)$ could be demonstrated by exclusive
$\bar{B}\to D^{(*)0} f_{0}(980)$ and $\bar{B}\to J/\Psi
f_{0}(980)$ processes and $f_{0}(980)$ could be reconstructed from
the decays $\bar{B}\to D^{(*)0} \pi^{+} \pi^{-}(KK)$ and
$\bar{B}\to J/\Psi \pi^{+} \pi^{-}(KK)$. The results could be as
the complement to the three-body decays of $D_{s}$ that already
indicate the existence of $s \bar{s}$ component .

It is known that the exclusive $b\to c$ decays are dominated by
the tree contributions and only $(V-A)\otimes (V-A)$ four-fermi
interactions need to be considered.
The difficulty in our calculations is how to determine the
involving wave functions which are sensitive to the
nonperturbative QCD effects and are universal objects. In $B$
meson case, one can fix it by $B\to PP$ processes, with $P$
corresponding to light pseudoscalars in which the wave functions
are defined in the frame of light-cone and have been derived from
QCD sum rule \cite{Ball}. As to the $D^{(*)0}(J/\Psi)$ wave
functions, we can call for the measured BRs of  color-suppressed
decays $\bar{B}\to D^{0} \pi^{0}$ \cite{BD} and $\bar{B}\to J/\Psi
K^{(*)}$ \cite{BJ}. However, it might be questionable to apply the
QCD approach for ordinary $ PP$ modes to $D^{(*)}(J/\Psi)$ decays
because they aren't light mesons anymore. In the heavy $b$ quark
limit, fortunately, the involved scales satisfy
$m_{b}>>m_{c}>>\bar{\Lambda}$ with $m_{b(c)}$ being the mass of
$b(c)$-quark and $\bar{\Lambda}=M_{B}-m_b$ so that the leading
power effects in terms of the expansions of $\bar{\Lambda}/m_{c}$
and $m_{c}/m_{b}$ could be taken as the criterion to estimate the
involving processes. We will see later that not only the obtained
BRs of $\bar{B}\to J/\Psi K^{*}$ but also their helicity
components of decay amplitudes
are  consistent with current experimental data. It will guarantee
that our predicted results on $f_{0}(980)$ productions of $B$
decays are reliable.

\begin{figure}
\includegraphics*[width=1.6
  in]{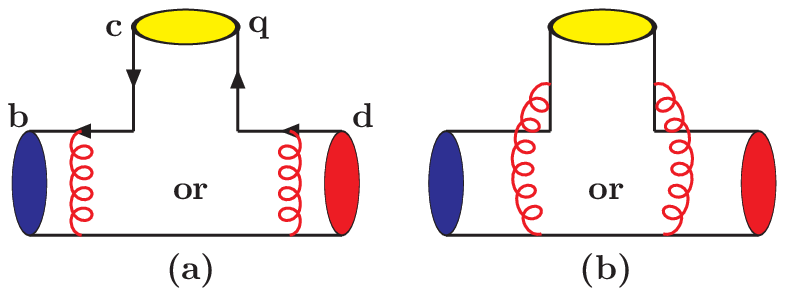} 
\includegraphics*[width=1.6
  in]{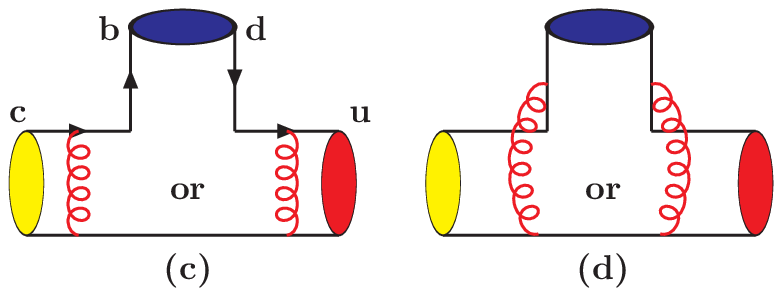}
\caption{Figures (a) and (b) illustrate the factorizable and
non-factorizable emission topologies, respectively, while
(c) and (d) correspond to the annihilation topologies.}
\label{topology}
\end{figure}

Since the hadronic transition matrix elements of penguin effects
in $\bar{B}\to J/\Psi M$, $M=K,\ K^{*}$, and $f_{0}(980)$, can be
related to tree ones, we  describe the effective Hamiltonian for
the $b\to c\bar{q} d$ transition as
\begin{eqnarray}
H_{{\rm eff}}&=&\frac{G_{F}}{\sqrt{2}}\sum_{q=u,c}V_{q}\left[
C_{1}(\mu ){\cal O}_{1}^{(q)}+C_{2}(\mu ){\cal O}
_{2}^{(q)}\right] \label{eff}
\end{eqnarray}
with ${\cal O}_{1}^{(q)} = \bar{d}_{\alpha} q_{\beta}
\bar{c}_{\beta} b_{\alpha}$ and  ${\cal O}_{2}^{(q)} =
\bar{d}_{\alpha} q_{\alpha} \bar{c}_{\beta} b_{\beta}$, where
$\bar{q}_{\alpha} q_{\beta}=\bar{q}_{\alpha} \gamma_{\mu}
(1-\gamma_{5}) q_{\beta}$, $\alpha(\beta)$ are the color indices,
$V_{q}=V_{qd}^{*}V_{cb}$ are the products of the CKM matrix
elements, and $C_{1,2}(\mu )$ are the Wilson coefficients (WCs)
\cite{BBL}. Conventionally, the effective WCs of
$a_{2}=C_{1}+C_{2}/N_{c}$ and $a_{1}=C_{2}+C_{1}/N_{c}$ with
$N_{c}=3$ being color number are more useful.  $q=u$ corresponds
to $\bar{B}\to D^{(*)0} M$ decays while $q=c$ stands for
$\bar{B}\to J/\Psi M$ decays.
According to the effective operators in Eq. (\ref{eff}), we find
that only emission topologies contribute to $\bar{B}\to J/\Psi
f_{0}(980)$, however, the decays of $\bar{B}\to D^{(*)0}
f_{0}(980)$ involve both emission and annihilation topologies. To
be more clear, the illustrated diagrams are displayed in Fig.
\ref{topology}. From the figure, we could see obviously that only
$n \bar{n}$ content has the contributions and the factorizable
emission parts, Fig. \ref{topology}(a), are only related to
the $\bar{B}\to f_{0}(980)$ form factor. We note that in the
color-suppressed processes the nonfactorizable effects, shown as
Fig. \ref{topology}(b) and (d), are important and should be
considered.

Regarding $f_{0}(980)$ as $q \bar{q}$ contents in $B$ decays, the
immediate question is how to write down the corresponding hadronic
structures and the associated wave functions for this $^{3}P_{0}$
state. What we know is that the spin structures of $f_{0}(980)$
should satisfy $\langle 0|\bar{q} \gamma_{\mu} q|
f_{0}(980)\rangle = 0$ and $\langle 0|\bar{q} q |f_{0}(980)\rangle
= m_{f_{0}} \tilde{f}$ in which $m_{f_{0}}(\tilde{f}\approx 0.18)$
\cite{DP} is the mass ( decay constant) of $f_{0}(980)$. In order
to satisfy these local current matrix elements, the light-cone
distribution amplitude for $f_{0}(980)$ should be given by
\begin{eqnarray}
\langle 0|\bar{q}(0)_{j} q(z)_{l}|f_{0}\rangle
&=&\frac{1}{\sqrt{2N_{c}}}\int^{1}_{0} dx e^{-ixP\cdot z} \Big\{
[\, \slash \hspace{-0.2cm} p\, ]_{lj} \Phi_{f_{0}}(x) \nonumber
\\ &&+m_{f}[{\bf 1}]_{lj} \Phi^{p}_{f_{0}}(x)\Big\} \label{daf}
\end{eqnarray}
where $\Phi^{(p)}_{f_{0}}(x)$ belong to twist-2(3) wave functions.
The charge parity indicates that
$\Phi_{f_{0}}(x)=-\Phi_{f_{0}}(1-x)$ and
$\Phi^{p}_{f_{0}}(x)=\Phi^{p}_{f_{0}}(1-x)$ \cite{CZ} so that
their normalizations are $\int^{1}_{0} dx\Phi_{f_0}(x)=0$ and
$\int^{1}_{0} dx\Phi^{p}_{f_0}(x)=\tilde{f}/2\sqrt{2N_{c}}$. As
usual, we adopt a good approximation that the light-cone wave
functions are expanded in Gegenbauer polynomials. Therefore, we
choose
\begin{eqnarray}
 \Phi^{p}_{f_{0}}(x)&=& \frac{\tilde{f}}{2\sqrt{2N_{c}}} \bigg\{
3(1-2x)^{2}+ G_{1}^{p} (1-2x)^{2} \nonumber
\\
&& \times \left[C^{3/2}_{2}(1-2x)-3 \right]+ G^{p}_{2}
C^{1/2}_4(1-2x)
\bigg\},\nonumber \\
\Phi_{f_{0}}(x)&=& {\tilde{f} \over 2\sqrt{2N_{c}}} G
\left[6x(1-x)C^{3/2}_{1}(1-2x)\right], \label{wavf}
\end{eqnarray}
where $C^{\lambda}_{n}$ are the Gegenbauer polynomials and the
values of coefficients $\{G\}$ haven't been determined yet from
the first principle QCD approach.

It has been shown that by the employ of hierarchy
$M_{B}>>M_{D^{(*)}}>>\bar{\Lambda}$, the $D^{(*)}$ meson
distribution amplitudes could be described by \cite{Li-P}
\begin{eqnarray}
\langle 0|\bar{d}(0)_{j} c(z)_{l}|D\rangle
&=&\frac{1}{\sqrt{2N_{c}}}\int^{1}_{0} dx e^{-ixP\cdot z}
\nonumber \\
&& \times \Big\{ [\, \slash \hspace{-0.2cm}
p\, +M_{D} ]_{lj} \gamma_{5} \Phi_{D}(x)\Big\}, \nonumber \\
\langle 0|\bar{d}(0)_{j} c(z)_{l}|D^{*}\rangle
&=&\frac{1}{\sqrt{2N_{c}}}\int^{1}_{0} dx e^{-ixP\cdot z}
\nonumber \\
&& \times \Big\{   [\, \slash \hspace{-0.2cm} p\, +M_{D^{*}}
]_{lj} \, \slash \hspace{-0.18 cm} \varepsilon \,
\Phi_{D^{*}}(x)\Big\},\label{dad}
\end{eqnarray}
where $\varepsilon_{\mu}$ is the polarization vector of $D^{*}$,
the normalizations of wave functions are taken as $\int^{1}_{0}
dx\Phi_{D^{(*)}}(x)=f_{D^{(*)}}/2\sqrt{2N_{c}}$ and $f_{D^{(*)}}$
are the corresponding decay constants. Although the decay
constants and wave functions of the $D^{*0}$ meson for longitudinal
and transverse polarizations are different generally, for
simplicity, in our estimations we assume that they are the same.
Since the hadronic structure of $B$  was studied before, the
explicit description can be found in Ref. \cite{Chen-Li}. In order
to fit the measured BR of $\bar{B}\to D^{0} \pi^0$, the
involved $D^{(*)}$ wave functions are modelled simply as
\cite{Li-P}
\begin{eqnarray}
 \Phi_{D^{(*)}}(x)&=&{3 \over \sqrt{2N_{c}}}
f_{D^{(*)}}x(1-x)[1+0.7 (1-2x)].
\end{eqnarray}
With the same guidance, we also apply the concept to the $J/\Psi$
case. As a detail discussion, one can refer Ref. \cite{YL}.

As mentioned before, due to a large energy transfer in heavy $B$
meson decays, we can utilize the factorization theorem, in which
decay amplitudes can be calculated by the convolution of hard
parts and wave functions \cite{LB,Li}, to describe the hadronic
effects. Although vector $D^{*0}$ and $J/\Psi$ mesons carry the
spin degrees of freedom, in the $\bar{B}\to D^{*0} (J/\Psi)
f_{0}(980)$ decays only longitudinal polarization is involved.
Expectably, the results should be similar to the $D^{0}f_{0}(980)$
mode. Hence, we only present the representative formulas for
$\bar{B}\to D^{0} f_{0}(980)$ at the amplitude level but give the
predicted BRs for all considered processes. From  Fig.
\ref{topology} and the effective interactions of Eq. (\ref{eff}),
the decay amplitude for $\bar{B}\to D^{0} f_{0}(980)$ is written
by
\begin{eqnarray*}
A_{\bar{n} n} &=& {\sin \phi_{s} \over \sqrt{2}} V_{u} \Big[f_{D
}{\cal F}_{e} +{\cal M}_{e}+f_{B} {\cal F}_{a}+{\cal M}_{a} \Big]
\end{eqnarray*}
where ${\cal F}_{e}({\cal M}_{e})$ and ${\cal F}_{a}({\cal
M}_{a})$ are the factorized (non-factorized) emission and
annihilation hard amplitudes, respectively. According to Eqs.
(\ref{daf}) and (\ref{dad}), the typical hard functions are
expressed as
\begin{eqnarray}
{\cal F}_{e}&=&\zeta \int_{0}^{1} dx_{1} dx_{3} \int_{0}^{\infty}
b_{1}db_{1}b_{3}db_{3}\Phi _{B}( x_{1},b_{1})
\nonumber\\
&&  \Big\{ \Big[(1+x_{3})\Phi_{f_{0}}(x_{3})
+r_{f}(1-2x_{3})\Phi^{p}_{f_{0}}(x_{3})\Big]{\cal
E}_{e}(t^{1}_{e}) \nonumber \\
&& +2r_{f}\Phi^{p}_{f_{0}}(x_3){\cal E}_{e}(t^{2}_{e}) \Big\},
\label{fe}\\
{\cal M}_{e}&=&2\zeta \int_{0}^{1} d[x] \int_{0}^{\infty}
b_{1}db_{1} b_{3}db_{3} \Phi _{B}( x_{1},b_{1})\Phi_{D}(x_{2})
\nonumber\\
&& \Big\{ \Big[-(x_{2}+x_{3}) \Phi_{f_{0}}(x_{3})
+r_{f}x_3\Phi^{p}_{f_{0}}(x_{3})\Big] {\cal
E}^{1}_{d}(t^{1}_{d})\nonumber \\
&&+\Big[(1-x_{2})\Phi_{f_{0}}(x_{3})-r_{f}x_{3}\Phi^{p}_{f_{0}}(x_3)\Big]{\cal
E}^{2}_{d}(t^{2}_{d}) \Big\}, \label{me}
\end{eqnarray}
with $\zeta=8\pi C_{F} M^{2}_{B}$, $r_{f}=m_{f}/M_{B}$, ${\cal
E}^{i}_{e}(t^{i}_{e})=$ $\alpha_{s}(t^{i}_{e})$ $a_{2}(t^{i}_{e})$
$Su_{B+f_{0}(980)}(t^{i}_{e})$ $h_{e}(\{x\},\{b\})$ and ${\cal
E}^{i}_{d}(t^{i}_{d})=\alpha_{s}(t^{i}_{d})(C_{2}(t^{i}_{d})/N_{c})$
$Su(t^{i}_{d})_{B+D+f_{0}(980)}$ $h_{d}(\{x\},\{b\})$.
$t^{1,2}_{e,d}$, $Su$ and $h_{e,d}$ denote the hard scales of $B$
decays, Sudakov factors and hard functions  arising from
the propagators of gluon and internal valence quark, respectively.
Their explicit expressions can be found in Ref. \cite{Chen-Li}.
With the same procedure,
the other hard functions can be also derived. 

So far, the still uncertain values are the $\{G\}$ parameters of
the $f_{0}(980)$ wave functions. By the identity of
$<0|\bar{q}\gamma_{\mu}q|V,T>=M_{V}f_{V}\varepsilon_{\mu}(T)$ for
the $V$-meson transverse polarization, we find that except the Dirac
matrices $\gamma_{\mu}$ and the associated polarization vector
$\varepsilon_{\mu}$, it is similar to the scalar meson case. Inspired
by the similarity, we adopt $\Phi^{p}_{f_{0}}(x)$ to be a
$\rho$-meson like wave function and take $G^{p}_{1}\approx 1.5$
and $G^{p}_{2}\approx 1.8$ \cite{Chen}. As to the value of $G$, we
use the corresponding value in $a_{0}(980)$  given by the
second reference of \cite{CZ} and get $G\approx 1.11$. By the
chosen values and using Eq. (\ref{fe}) with excluding WC of
$a_{2}$, we immediately get the $\bar{B}\to f_{0}(980)$ form
factor to be 0.38. Is it a reasonable value? In order to
investigate that the obtained value is proper, we employ the
relationship $F^{B\to f_{0}(980)}\sim (M_{D_{s}}/M_{B})^{1/2}
F^{D_{s}\to f_{0}(980)}$, which comes from the heavy quark symmetry
limit \cite{HQ}, as a test. According to the calculation of Ref.
\cite{DGNPT}, we know $F^{D_{s}\to f_{0}(980)}\approx 0.6$; and
then, we have $F^{B\to f_{0}(980)}\sim 0.36$. Clearly, it is quite
close to what we obtain. Hence, with the taken values of
parameters, the magnitudes of hard functions are given in Table
\ref{tablehf}. We note that the complex values come from the
on-shell internal quark and all of hard functions are the same in
order of magnitude.
\begin{table}[htb]
\caption{Hard functions (in units of $10^{-2}$) for $\bar{B}\to
D^{0} f_{0}(980)$ decay with 
$\tilde{f}=0.18$, $f_{D}=0.2$ GeV, $G=1.11$, $G^{p}_{1}=1.5$ and
$G^{p}_{2}=1.8$. }\label{tablehf}
\begin{ruledtabular}
\begin{tabular}{ccccc}
Amp. &  $ \ \ {\cal F}_{e} \  \ $ & $\  \ {\cal M}_{e}\ \ $ & $\ \
{\cal
F}_{a} \ \ $ & $\ \ {\cal M}_{a}\ \ $  \\
\hline $D^{0}f_{0}(980)$ & $-5.95$  &  $-2.66+i1.56$ &
$1.83-i3.60$ & $0.20+i1.12$
\end{tabular}
\end{ruledtabular}
\end{table}

\begin{figure}
\includegraphics*[width=2.6
  in]{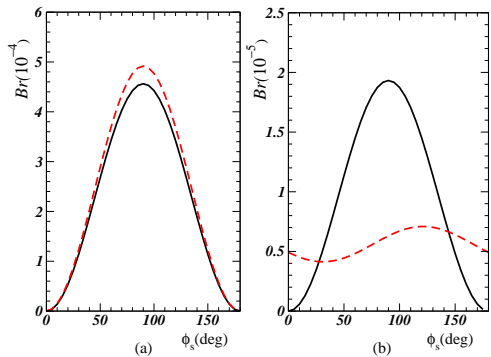} 
\caption{ BRs as a function of angle $\phi_{s}$. (a) the solid
(dashed) lines are for $\bar{B}\to D^{0}(D^{*0}) f_{0}(980)$
decays while (b) they express $\bar{B}\to J/\Psi f_{0}(980)$ and
$B^{+}\to K^{+} f_{0}(980)$ decays. }\label{fig}
\end{figure}

One challenging question is that how reliable our results are. In
order to investigate this point, besides the $\bar{B}\to D^{(*)0}
(J/\Psi) f_{0}(980)$ decays, we also calculate $\bar{B}\to D^{0}
\pi,\ J/\Psi (K, K^{*})$ and $\bar{B}\to f_{0}(980)K^{+}$
processes. All of them are already measured at B factories
\cite{BelleBr,BaBarBr}. Due to the calculations and formalisms
being similar to $D^{0}f_{0}(980)$, we directly present the
predicted BRs in Table \ref{tablebr} by taking $\phi_{s}=45^{0}$,
$f_{D^{*}}=0.22$ GeV, $f_{J/\Psi}=0.405$ GeV and the same taken
values of Table \ref{tablehf}. As to the $J/\Psi$ wave functions,
we model it as $\Phi_{J/\Psi}(x)=f_{J/\Psi}
[30x^{2}(1-x)^{2}]/2\sqrt{2N_{c}}$. The BRs of charged $B^{+}\to
J/\Psi M^{+}$ modes can be obtained from neutral modes by using
$Br(\bar{B}^{0}\to J/\Psi M^{0})\tau_{B^+}/\tau_{B^0}$. Hence,
 from the Table \ref{tablebr}, we clearly see that our predictions
are consistent with experimental data.
\begin{table}[h]
\caption{BRs (in units of $10^{-4}$) with $\phi_{s}=45^{0}$,
$f_{D^{*}}=0.22$, $f_{J/\Psi}=0.405$ GeV and the same taken values
of Table \ref{tablehf}.} \label{tablebr}
\begin{ruledtabular}
\begin{tabular}{cccc}
Mode &   Belle \cite{BelleBr} & BaBar \cite{BaBarBr} & This  \vspace{-0.2cm}  \\
 &  &  & work \\
\hline $D^{0}f_{0}(980)$ &   &   &
$2.28$ \\ \hline 
       $D^{*0}f_{0}(980)$ &   &  & $2.46$\\ \hline 
       $J/\Psi f_{0}(980)$ & &  & $0.10$ \\ \hline 
       $K^{+} f_{0}(980)$ &  &  &    \\  
        $f_{0}\to \pi^{+} \pi^{-}$ & $(9.6^{+2.5+1.5+3.4}_{-2.3-1.5-0.8})\cdot 10^{-2}$ &  &  $0.02$\\ \hline 
       $D^{0}\pi^{0}$ & $3.1\pm 0.4 \pm 0.5$  &  $2.89\pm 0.29 \pm 0.38$ & $2.60$ \\ \hline  
       $J/\Psi K^{0}$ & $7.9\pm 0.4 \pm 0.9$  & $8.3\pm 0.4 \pm 0.5$ & $8.3$ \\ \hline 
       $J/\Psi K^{*0}$ & $12.9\pm0.5 \pm 1.3$  & $12.4\pm 0.5 \pm 0.9$&  $13.37$
\end{tabular}
\end{ruledtabular}
\end{table}
Moreover, it is worthful to mention that in addition to the $BR$
of $\bar{B}\to J/\Psi K^*$ decay, the squared helicity amplitudes
$|A_{0}|^{2}$, $|A_{\parallel}|^{2}$ and $|A_{\perp}|^{2}$ with
the normalization of $|A_{0}|^{2}+|A_{\parallel}|^{2}+
|A_{\perp}|^{2}=1$ \cite{CKL} are also given as $0.59$, $0.24$ and
$0.17$, respectively. They are all comparable with the measured
values $0.60\pm0.05 (0.60\pm 0.04)$, $0.21\pm 0.08(0.24\pm0.04)$
and $0.19\pm0.06 (0.16\pm 0.03)$ of Belle (BaBar)
\cite{BelleBr,BaBarBr}. In order to further understand the
dependence of the effects of $n \bar{n}$ content, the BRs as a
function of mixing angle $\phi_{s}$ are shown in Fig. \ref{fig}.
We note that with including the twist-2 wave function for
$f_{0}(980)$, our previous result of $B^{+}\to K^{+} f_{0}(980)$
in the small $\phi_{s}$ region \cite{Chen} becomes insensitive to
$\phi_{s}$.

The subsequent question is how to search the events for
$\bar{B}\to D^{(*)0}f_{0}(980)$ and $\bar{B}\to J/\Psi f_{0}(980)$
decays. From particle data group of Ref. \cite{BJ}, we know that
$f_{0}(980)$ mainly decays to $\pi \pi$ and $K K$  and
$R=\Gamma(\pi\pi)/(\Gamma(\pi\pi)+\Gamma(KK))\sim 0.68$.
Therefore, we suggest that the candidates could be found in
$\bar{B}\to D^{(*)0}(J/\Psi)\ \pi \pi\ (KK)$ three-body decay
samples. For an illustration, according to the values of Table
\ref{tablebr}, we can estimate that the BR product of
$Br(\bar{B}\to D^{0}f_{0}(980))\times Br(f_{0}(980)\to \pi^+
\pi^-)\approx 1.0\times 10^{-4}$ with $Br(f_{0}(980)\to \pi^+
\pi^-)= 2R/3$. The result is consistent with the measured value of
$(8.0 \pm 0.6\pm 1.5) \times 10^{-4}$ for $\bar{B}\to D^{0}
\pi^{+} \pi^{-} $ decay while that of $\bar{B}\to D^{0} \rho^{0}$  is
determined to be $(2.9\pm 1.0 \pm 0.4) \times 10^{-4}$
\cite{Belle-conf}.


We have investigated the possibility to extract the existence of
$n \bar{n}$ component of $f_{0}(980)$ in terms of $\bar{B}\to
D^{(*)0} f_{0}(980)$ and $\bar{B}\to J/\Psi f_{0}(980)$ decays.
Based on the comparable values between the BRs of $\bar{B}\to
D^{0}\pi^{0}$ and $\bar{B}\to J/\Psi M$ decays and current
experimental data, our predictions on the BRs of $\bar{B}\to
D^{(*)0}(J/\Psi) f_{0}(980)$ decays are reliable
and can be tested at $B$ factories.\\

\noindent {\bf Acknowledgments:}

The author would like to thank H.N. Li and  H.Y. Cheng for their
useful discussions. This work was supported in part by the
National Science Council of the Republic of China under Grant No.
NSC-91-2112-M-001-053. \\


\end{document}